\documentclass[aps,pra,superscriptaddress,amsmath,amssymb,preprintnumbers,showpacs,twocolumn]{revtex4-1}
\usepackage{amssymb}
\usepackage{graphicx}
\usepackage{bm}
\usepackage{color}
\usepackage{subfigure}

\newcommand{\Q}{\dot{Q}}

\begin{document}

\title{Quantum thermal machines with single nonequilibrium environments}

\author{Bruno Leggio}
\affiliation{Laboratoire Charles Coulomb, UMR 5221 Universit\'{e} de Montpellier and CNRS, F- 34095 Montpellier, France}

\author{Bruno Bellomo}
\affiliation{Laboratoire Charles Coulomb, UMR 5221 Universit\'{e} de Montpellier and CNRS, F- 34095 Montpellier, France}

\author{Mauro Antezza}
\affiliation{Laboratoire Charles Coulomb, UMR 5221 Universit\'{e} de Montpellier and CNRS, F- 34095 Montpellier, France}
\affiliation{Institut Universitaire de France, 103 Boulevard Saint-Michel, F-75005 Paris, France}

\newcommand{\ket}[1]{\displaystyle{|#1\rangle}}
\newcommand{\bra}[1]{\displaystyle{\langle #1|}}

\date{\today}

\begin{abstract}
We propose a scheme for a quantum thermal machine made by atoms interacting with a single non-equilibrium electromagnetic field. The field is produced by a simple configuration of macroscopic objects held at thermal equilibrium at different temperatures. We show that these machines can deliver all thermodynamic tasks (cooling, heating and population inversion), and this by establishing quantum coherence with the body on which they act. Remarkably, this system allows to reach efficiencies at maximum power very close to the Carnot limit, much more than in existing models. Our findings offer a new paradigm for efficient quantum energy flux management, and can be relevant for both experimental and technological purposes.
\end{abstract}

\pacs{}

\maketitle
\section{Introduction}
Recent years have seen an uprising interest in thermodynamics at atomic scale \cite{GemmerBook, Blicke2012, Brunner2012, Horodecki2013} due to the latest-generation manipulation of few, if not single, elementary quantum systems \cite{Blicke2012, Haroche2013}. In particular, out-of-equilibrium thermodynamics of quantum systems represents one of the most active research areas in the field \cite{Esposito2009, Deffner2011, Leggio2013a, Leggio2013b,Abah2014}. In this context, triggered by vast technological outcomes \cite{Scully2010, Haenggi2009}, the concept of quantum absorption thermal machine \cite{Scovil1959} has been reintroduced \cite{Linden2010, Levy2012, Correa2014, Venturelli2013, Correa2013, Brunner2014}. These machines are particularly convenient since they function without external work, extracting heat from thermal reservoirs through single atomic transitions to provide thermodynamic tasks (e.g., refrigeration).

Nonetheless, fundamental issues remain unsolved. A first one is the connection a single atomic transition to given thermal reservoirs, posing serious obstacles to practical realizations of such machines. A second, more theoretical issue concerns the role of quantumness. Indeed, in typical models quantum features are not required \cite{Scovil1959,Correa2013}, and only recently the advantages of quantum properties in thermal reservoirs have been pointed out \cite{Correa2014}. The role of quantum features in the machines itself is debated \cite{Correa2013, Brunner2014}, so that the advantages of quantum machines over standard ones remains partially unclear.
\begin{figure}[t!]
        \begin{center}
       \subfigure[]{\includegraphics[width=220pt]{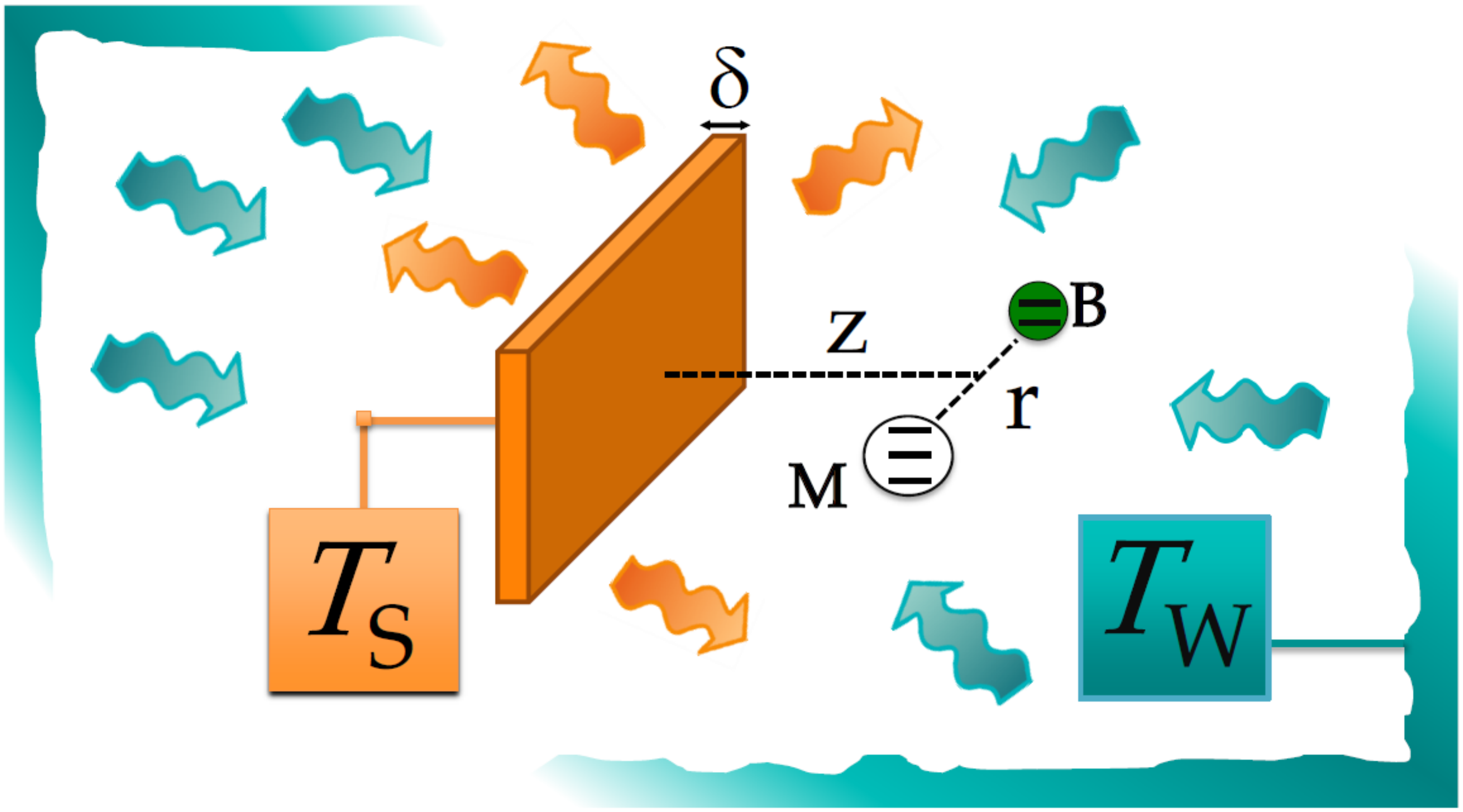} \label{model}}
        \subfigure[]{\includegraphics[width=120pt]{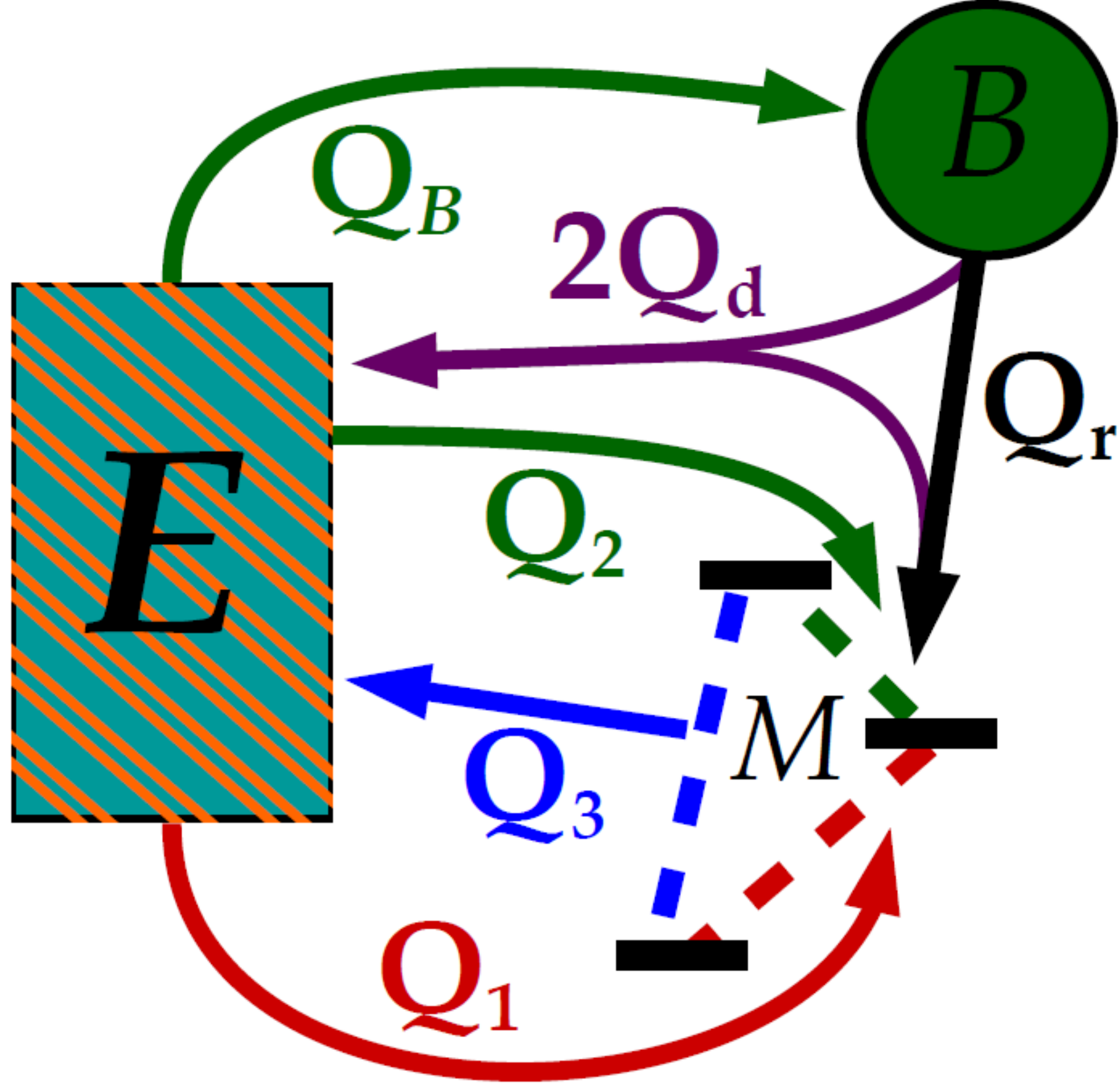} \label{scheme}}
        \caption{(a) A slab of thickness $\delta$ at temperature $T_S$ is placed in the blackbody radiation of some walls at temperature $T_W$. Two atoms are placed in the resulting OTE electromagnetic field, at a distance $z$ from the slab and at a distance $r$ from each other. (b) Stationary heat fluxes between the OTE environment $E$ and each atomic transition. The OTE field also mediates for an effective atomic interaction producing an energy exchange $Q_r$ between resonant atomic transitions. Each flux contribution corresponds to a term in the atomic master equation \eqref{METQ}.}\label{system}
  \end{center}
\end{figure}

In this paper we address both of these open problems by introducing a new quantum thermal machine setting, based on an out-of-thermal-equilibrium (OTE) electromagnetic bath naturally (i) coupling to each single atomic transitions, and (ii) creating quantum features in the machine. The field is produced by macroscopic objects, and acts on each atomic transition as a different thermal bath at an effective temperature, hence providing all the elements needed for quantum absorption tasks.

This paper is structured as follows: the physical system is introduced in Section \ref{system}, along with the master equation governing its dynamics, while Section \ref{thermodynamics} is devoted to the introduction of thermodynamic quantities characterising the heat exchanges happening between atoms and field. In Section \ref{tasks} the first part of the results is given, concerning the action of the machines, the different tasks it can produce and their intrinsic quantum origin. The second part of the results of this work about the machine efficiency and its Carnot limit is given in Section \ref{efficiency}. Finally, remarks and conclusions are drawn in Section \ref{conclusions}.

\section{Physical system}\label{system} The setup of this paper is schematically depicted in Fig. \ref{system}, where a slab of thickness $\delta$ at temperature $T_S$ is placed in the blackbody radiation emitted by some walls at temperature $T_W\neq T_S$. The total electromagnetic field embedding the space between the slab and the walls is therefore given by the sum of four contribution: the direct blackbody radiation of the walls, the radiation emitted naturally by the slab and the walls' radiation after being either reflected or transmitted by the slab. Such an OTE field has been studied in the context of Casimir-Lifshitz force and heat transfer \cite{Antezza2004, Antezza2005, Antezza2006, Messina2011}, where its properties have been characterised in terms of the field correlators through a scattering matrix approach. The slab and the walls, macroscopic objects, are here the only ones directly connected to thermal baths.
In addition, a three-level atom $M$ (machine) and a two-level atom $B$ (target body) are placed at the same distance $z$ from the surface of the slab and spatially separated by a distance $r$. The atomic open system involves then four transitions: the body transition labeled as $B$ and the three machine transitions labeled as $1,2,3$. Transition $1$ connects the two lowest-lying energy eigenstates (red transition in Fig. \ref{scheme}) and transition $2$ the two highest ones (green transition in Fig. \ref{scheme}). The OTE field interacts with them through the Hamiltonian $H_I=-\sum_i\mathbf{d}_i \cdot \mathbf{E}(\mathbf{R}_i)$, where $\mathbf{d}_i$ is the dipole moment of the $i$-th transition of the atomic system and $\mathbf{E}(\mathbf{R}_i)$ is the electromagnetic field at its position $\mathbf{R}_i$. The total Hamiltonian of the system is
\begin{equation}
H_{tot}=H_M+H_B+H_{\mathrm{field}}+H_I,
\end{equation}
where $H_{\mathrm{field}}$ is the Hamiltonian of the OTE field. In the following we will not need the explicit expression of $H_{\mathrm{field}}$ since only the field correlations will enter the master equation describing the dynamics of the atoms. The free atomic Hamiltonians $H_M$ and $H_B$ have expressions
\begin{eqnarray}
H_B&=&\big(\omega_B +\Delta S(\omega_B)\big)\sigma^{\dag}\sigma=\widetilde{\omega}_B\sigma^{\dag}\sigma,\label{HQ}\\
H_M&=&\big(\omega_1 +\Delta S(\omega_1)+S^-(\omega_2)-S^-(\omega_3)\big)\kappa_1^{\dag}\kappa_1\nonumber \\
&+&\big(\omega_3 +\Delta S(\omega_3)+S^+(\omega_2)-S^-(\omega_1)\big)\kappa_3^{\dag}\kappa_3\nonumber\\
&=&\widetilde{\omega}_1\kappa_1^{\dag}\kappa_1+\widetilde{\omega}_3\kappa_3^{\dag}\kappa_3,\label{HT}
\end{eqnarray}
being $\sigma$ ($\sigma^{\dag}$) the lowering (raising) operator of the body $B$ and $\kappa_n$, $\kappa_n^{\dag}$ ($n=1,2,3$) the lowering and raising operators of $M$.
$S^{\pm}(\omega_i)$ here represents a shift of the energy of each level in the $i-$th transition due to the local interaction with the field and $\Delta S(\omega)=S^+(\omega)-S^-(\omega)$. In the third equality the renormalised transition frequencies $\widetilde{\omega}_i$ for $M$ and $B$ have been introduced to account for the effects of the shifts $S^{\pm}(\omega_i)$. Throughout this work we will always assume the physical consequences of such frequency renormalisation to be negligible, such that $\widetilde{\omega}_i=\omega_i\,\,\forall i$. This assumption has been fully confirmed by extended numerical simulations, having always detected the relative error introduced by neglecting these shifts to be less than $1\%$.

It is worth stressing here that, differently from previous works on atomic-scale thermal machines \cite{Linden2010, Levy2012, Correa2014, Correa2013, Brunner2014}, each atomic transition interacts here with \textit{the same} electromagnetic field, which embeds all the space where the atoms are placed. As we will show in what follows, there is then no need to conceive different environments, each interacting with a single atomic transition: a single non-equilibrium electromagnetic field is here able to produce all the physics needed for quantum thermodynamic tasks.
\subsection{The master equation}
In \cite{Bellomo2013} the master equation (ME) for two emitters in such a field has been derived under the Markovian limit as
\begin{equation}\label{METQ}
\frac{d\rho}{d t}=-\frac{i}{\hbar}\big[H_T,\rho\big]+D_{B}(\rho)+\sum_{n=1}^3D_n(\rho)+D_d(\rho),
\end{equation}
where $H_T=H_M+H_B+H_{MB}$. $H_{MB}=\hbar\Lambda(\omega_B)(\sigma^{\dag}\kappa_r+\sigma \kappa_r^{\dag})$ is an effective \textit{field-mediated} dipole interaction coupling resonant atomic transitions. Here we assume $B$ and $M$ to be resonant through transitions at frequency $\omega_B$, and all their dipoles to have the same magnitude and to lie along the line joining the two atoms, and oriented from $B$ to $M$. $\hbar\Lambda(\omega)$ is the effective interaction strength and $\sigma$ ($\kappa_r$) is the lowering operator of the body (of the resonant transition of the machine). $H_{MB}$ originates from the correlations of the fluctuations of atomic dipoles due to the common field.

The derivation of the master equation \eqref{METQ} has been performed under the Markovian and rotating wave approximations. It involves the average photon number $n(\omega, T)=1/\big(e^{\hbar \omega/k_B T}-1\big)$ at frequency $\omega$ and temperature $T$ and the two functions $\alpha_{W(S)}$ which encompass all the properties of the environment, such as the dielectric properties of the slab and the correlation functions of the field. For their explicit expressions we refer the interested reader to \cite{Bellomo2013}.
The dissipative effects due to the atom-field coupling are accounted for by the dissipators $D_k$ with expressions
\begin{eqnarray}
D_B(\rho)&=&\Gamma_B^+(\omega_B)\Big(\sigma\rho\sigma^{\dag}-\frac{1}{2}\big\{\sigma^{\dag}\sigma,\rho\big\}\Big)\nonumber \\
&+&\Gamma_B^-(\omega_B)\Big(\sigma^{\dag}\rho\sigma-\frac{1}{2}\big\{\sigma\sigma^{\dag},\rho\big\}\Big),\label{DQ}
\end{eqnarray}
\begin{eqnarray}
D_n(\rho)&=&\Gamma_n^+(\omega_n)\Big(\kappa_n\rho\kappa_n^{\dag}-\frac{1}{2}\big\{\kappa_n^{\dag}\kappa_n,\rho\big\}\Big)\nonumber \\
&+&\Gamma_n^-(\omega_n)\Big(\kappa_n^{\dag}\rho\kappa_n-\frac{1}{2}\big\{\kappa_n\kappa_n^{\dag},\rho\big\}\Big),\label{DT}
\end{eqnarray}
\begin{eqnarray}\label{Dnonloc}
D_d(\rho)&=&\Gamma_d^+(\omega_B)\Big(\kappa_{r}\rho\sigma^{\dag}-\frac{1}{2}\big\{\sigma^{\dag}\kappa_{r},\rho\big\}\Big)\nonumber \\
&+&\Gamma_d^-(\omega_B)\Big(\kappa_{r}^{\dag}\rho\sigma-\frac{1}{2}\big\{\sigma\kappa_{r}^{\dag},\rho\big\}\Big)+h.c.\label{DTQ}
\end{eqnarray}
where $\omega_d=\omega_B$.
One recognises standard local dissipation terms ($D_B$ and $D_n$), each associated to the degrees of freedom of a well-identified atom, and non-local dissipation ($D_d$) which describes energy exchanges at frequency $\omega_B$ of the atomic system \textit{as a whole} with its OTE environment, not separable in machine or body contributions, its action involving degrees of freedom of both atoms in a symmetric way.
The parameters $\Gamma_i^{\pm}(\omega_i)$ (the rates of the dissipative processes of absorption and emission of photons through local or non-local interactions) depend on local or non-local correlations of the field in the atomic positions, which in turn are functions of the temperatures $T_S$ and $T_W$ and the dielectric properties of the slab $S$ as
\begin{eqnarray}
\frac{\Gamma^+_i(\omega)}{\Gamma^0_i(\omega)}&=&\big[1+n(\omega,T_W)\big]\alpha_W^i(\omega)\nonumber \\
&+&\big[1+n(\omega,T_S)\big]\alpha_S^i(\omega),\\
\frac{\Gamma^-_i(\omega)}{\Gamma^0_i(\omega)}&=&n(\omega,T_W)\alpha_W^i(\omega)^*+n(\omega,T_S)\alpha_S^i(\omega)^*,
\end{eqnarray}
where $\Gamma^0_i(\omega)=|\mathbf{d}_i|^2\omega^3/(3\hbar \pi \varepsilon_0 c^3)$, for $i=1,2,3,B$, is the vacuum spontaneous emission rate of the $i$-th atomic transition having a dipole moment $\mathbf{d}_i$, and $\Gamma^0_d(\omega)=\sqrt{\Gamma^0_B(\omega)\Gamma_{r}^0(\omega)}$. Thanks to the functional dependence of these parameters on the frequency and on the position of the atom, and to the critical behaviour shown in correspondence to the resonance frequency $\omega_S$ of the slab material, thermodynamic tasks become achievable. To simplify the notation, in the rest of this work the explicit $\omega$-dependence in all the $\Gamma$s will be omitted.

\section{Thermodynamics of the system}\label{thermodynamics}
After having introduced all the dynamic effects characterizing the atomic system, we want in this Section to introduce some quantities which will characterize the machine tasks and functioning.
\subsection{Environmental and population temperatures} In order to describe the machine thermodynamics, it is convenient to introduce two kind of temperatures. A first one characterizes the action of the field on the atoms: it has been shown \cite{Bellomo2012} that the atom-field interaction can be effectively rewritten as if each atomic transition felt a local \textit{equilibrium} environment whose temperature depends on the transition frequency, on the properties of the slab and on the slab-atom distance $z$. These effective \emph{environmental temperatures} depend on the rates $\Gamma_n^{\pm}$ as
\begin{equation}\label{Tn}
T_n=T(\omega_n)=\frac{\hbar \omega_n}{k_B \ln(\Gamma_n^+/\Gamma_n^-)},
\end{equation}
with $n=1,2,3,B,d$. It is important to stress here that, despite these effective environments can be characterised by a temperature, their spectra are not simply blackbody spectra as they have their own transition-dependent Purcell factor \cite{Bellomo2012}.\linebreak
In this framework we study thermodynamic effects of \textit{stationary} heat fluxes between $M$ and $B$, mediated and sustained by the OTE environment. To characterize the effects of these fluxes a second kind of temperature has to be introduced. Indeed, as much as the environmental temperatures characterise the thermodynamics of the OTE field, we need a second parameter to describe the energetics of atoms. In particular, atoms exchange energy under the form of heat with their surroundings by emitting photons through one of their transitions. This means that the possibility of such heat exchanges is related to the distribution of population in each atomic level. Note that, from the very definition of $T_n$, the environmental temperature depends on how the field tends to distribute atomic population in each pair of levels, due to the presence of the ratio $\Gamma_n^+/\Gamma_n^-$. A transition is therefore in equilibrium with its effective local environment if and only if its two levels $|a\rangle$ and $|b\rangle$ are populated such that $p_a/p_b=\Gamma^+/\Gamma^-$. If not, the field and the atom will exchange heat along such a transition until such a ratio is reached. This suggests to introduce a second temperature, hereby referred to as \textit{population temperature}, which for a transition of frequency $\omega_n$ ($n=1,2,3,B$) is defined as
\begin{equation}\label{thetai}
\theta_n=\frac{\hbar \omega_n}{k_B \ln(p^a_n/p^b_n)},
\end{equation}
$p^a_n$ ($p^b_n$) being the stationary population of the ground (excited) state of the $n$-th transition.
The result of a stationary thermodynamic task on the body, be it refrigeration, heating or population inversion, is then to modify its population temperature $\theta_B$.
\subsection{Heat fluxes}
The condition $T_i=\theta_i$ is satisfied only if detailed balance ($p^a_i/p^b_i=\Gamma_i^+/\Gamma_i^-$) holds. It can be proven that detailed balance can be broken in a three level atom in OTE fields. As a consequence the machine $M$ produces non-zero stationary heat fluxes with $B$ and the field environment, one for each dissipative process $D_n$ in the ME \eqref{METQ}. These fluxes, following the standard approach in the framework of Markovian open quantum systems \cite{BreuerBook}, are given as $\Q_n=\mathrm{Tr}\big[H_{at} D_n \rho\big]$, where $\rho$ is here the stationary atomic state and $H_{at}$ is a suitable atomic Hamiltonian which can be $H_M$, $H_B$ or $H_M+H_B$ depending on which part of the atomic system the heat flows into. Note that this definition implies an outgoing heat flux to be negative.

Following their definition, these heat fluxes depend both on the field properties (through the structure of the dissipators $D_n$) and on the properties of the atoms through their stationary state. This dependence, for the local dissipators, can be put under the very clear thermodynamic form
\begin{equation}\label{fluxtemp}
\dot{Q}_n=K_n\left(e^{\frac{\hbar \omega_n}{k_B \theta_n}}-e^{\frac{\hbar \omega_n}{k_B T_n}}\right)\simeq C_n(\theta_n)\big(T_n-\theta_n\big),
\end{equation}
where $K_n>0$, $C_n(\theta_n)$ is a positive function of $\theta_n$ (and of other parameters such as the frequency of the transition) and the second approximated equality holds in the limit $\theta_n\simeq T_n$. Equation \eqref{fluxtemp} shows that the direction of heat flowing is uniquely determined by the sign of the difference $T_n-\theta_n$, matching the thermodynamic expectation that heat flows naturally from the hotter to the colder body and strengthening the physical meaning of $\theta_n$.

There being no time-dependence in the Hamiltonian of the model, the first law of thermodynamics at stationarity for the total atomic system comprises only heat terms and assumes the form
\begin{equation}\label{1stlaw}
\Q_B+\sum_{n=1}^3\Q_n+\Q_d=0.
\end{equation}
In addition, energy is exchanged between the machine and the body thanks to their field-induced interaction $H_{MB}$. In Appendix A, following the general scheme developed in \cite{Weimer2008}, we show such an exchange to be under the form of heat. Seen by $M$ such a flux is $\Q_r=-i\mathrm{Tr}\big(H_M[H_{MB},\rho_s]\big)/\hbar$ while as expected $B$ sees the flux $-\Q_r$.
By introducing the explicit expressions for $H_M$ and $H_{MB}$, one can obtain a particularly simple form for $\Q_r$ as
\begin{equation}\label{Qrcorr}
\Q_r=2\hbar\omega_B\Lambda(\omega_B)\langle \sigma^{\dag}\kappa_r \rangle_{-},
\end{equation}
where $\langle \sigma^{\dag}\kappa_r \rangle_-=\frac{i}{2}\langle\sigma^{\dag}\kappa_r-\sigma\kappa_r^{\dag}\rangle$. In an analogous way, by employing Eq. \eqref{Dnonloc}, one can evaluate the change in internal energy of $M$ due to the non-local heat flux exchanged by the atomic system with the OTE environment, given by $\Q_d=\mathrm{Tr}\big[H_M D_d \rho\big]$. It is
\begin{equation}\label{Qdexpl}
\Q_d=-\hbar\omega_B\mathrm{Re}\Big\{ \langle \sigma\kappa_r^{\dag} \rangle\big[\Gamma_d^+-\left(\Gamma_d^-\right)^*\big]\Big\}.
\end{equation}
Finally, the change in the internal energy of $B$ due to the same effect, $\mathrm{Tr}\big[H_B D_d \rho\big]$, is given by same expression \eqref{Qdexpl}.

Fig. \ref{scheme} shows the full scheme of such heat fluxes for a particular configuration of the system.
The two levels of $B$ will be labeled here as $|g\rangle$ and $|e\rangle$. Despite the two-level assumption might seem specific, it has been shown in various contexts \cite{Brunner2012, DeLiberato2011} that quantum thermal machines only couple to some effective two-level subspaces in the Hilbert space of the body they are working on. A two-level system is therefore the fundamental building block of the functioning of quantum thermodynamic tasks.

\section{Coherence-driven machine tasks}\label{tasks} The main result of this paper is the possibility to drive the temperature $\theta_B$ of the body outside of the range defined by the external reservoirs at $T_W$ and $T_S$. The body, without the effect of the machine, would thermalise at the local environmental temperature ($\theta_B=T_B$), corresponding to $p_e/p_g=\Gamma_B^-/\Gamma_B^+$. This temperature is necessarily constrained within the range $[T_W,T_S]$ \cite{Bellomo2012}.\linebreak
Due to the particular form of the master equation \eqref{METQ}, in which all collective atomic terms involve only resonant atomic transitions, in the non-resonant subspace the collective atomic state will be diagonal in the eigenbasis of $H_B+H_M$. This is due to the fact that local dissipation in Eq. \eqref{METQ} of Section \ref{system} induces a thermalisation with respect to the free atomic Hamiltonians. On the other hand, in the resonant atomic subspace of the eigenbasis of $H_B+H_M$ spanned by the states $|g\rangle, |e\rangle$ of $B$ and the two states $|0_r\rangle, |1_r\rangle$ of the transition of $M$ at frequency $\omega_B$, the most general form of the atomic stationary state is
\begin{equation}\label{rhoX}
\begin{split}\begin{pmatrix}
 p_{e1_r} & 0 & 0 & 0\\
 0 & p_{e0_r} & c_r & 0\\
 0 & c_r^* & p_{g1_r} & 0\\
 0 & 0 & 0 & p_{g0_r}\\
\end{pmatrix}.
\end{split}\end{equation}
A coherence $c_r$ is present in the decoupled basis between the two atomic states $|g1_r\rangle$ and $|e0_r\rangle$ having the same energy.

\begin{figure}[t!]
\begin{center}
\includegraphics[width=220pt]{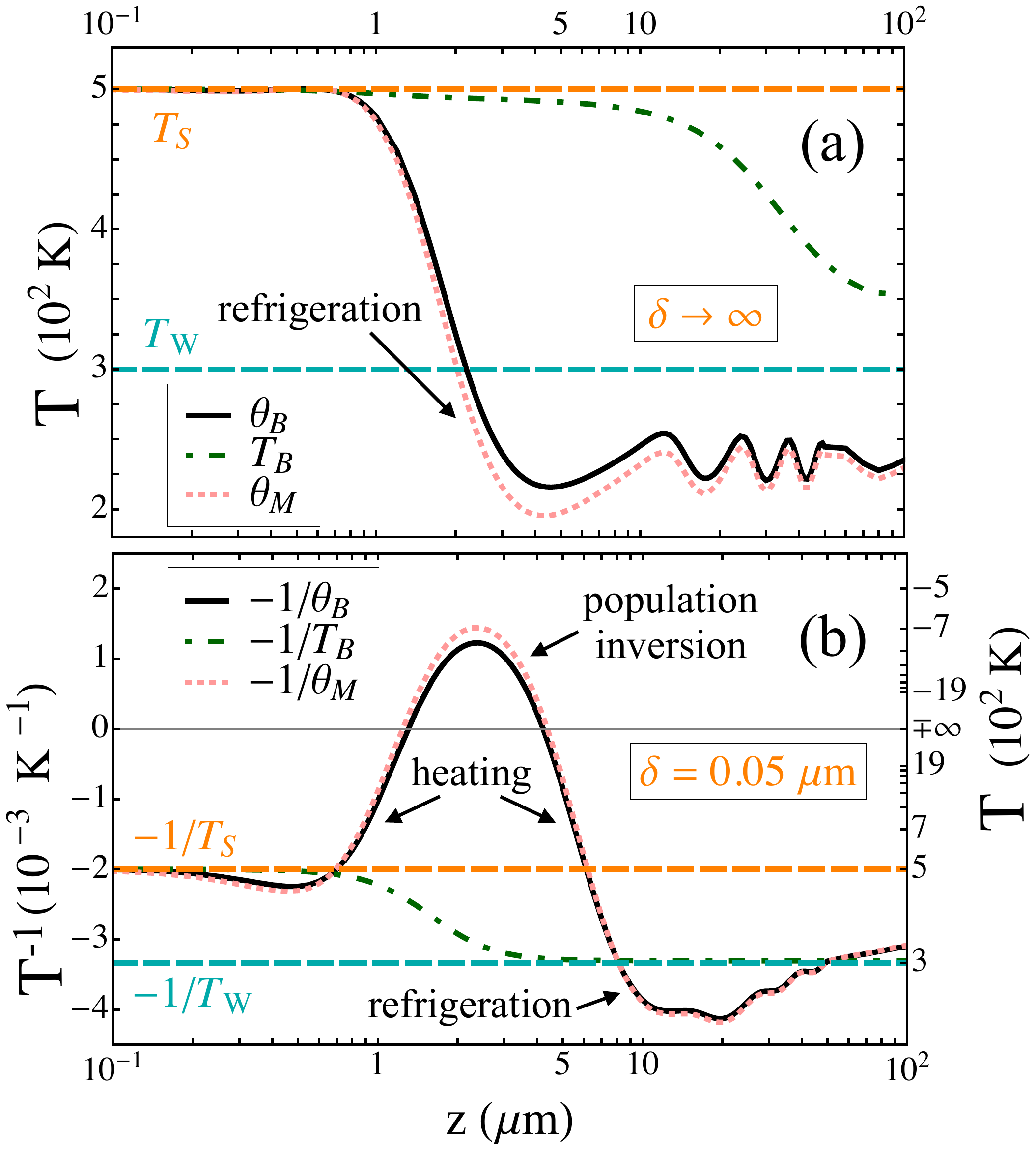}
\end{center}
\caption{Stationary temperature $\theta_B$ of the body (solid black line), machine resonant temperature $\theta_M$ in the absence of $B$ (dotted pink line) and local-environment temperature $T_B$ felt by the body (dot-dashed green line) versus $z$. The slab is made of sapphire and kept at $T_S=500\,\mathrm{K}$ while $T_W=300\,\mathrm{K}$. The machine transition frequencies are $\omega_1=0.9\,\omega_S$, $\omega_B=\omega_2=0.1\,\omega_S$ and $\omega_3=\omega_S$, $\omega_S=0.81\cdot10^{14}\,\mathrm{rad}s^{-1}$ being the first resonance frequency of sapphire (the optical data for the dielectric permittivity of the slab material are taken from \cite{PalikBook}). The two atoms are placed at a distance $r=1\,\mu\mathrm{m}$ from each other.
Panel (a): numerical data for a semi-infinite slab. Light and strong refrigeration are achieved in this configuration.
Panel (b): same quantities for a slab of finite thickness. The plotted functions are in this case $-1/\theta_B$, $-1/\theta_M$ and $-1/T_B$ (left vertical scale), with the same color code as before. The population inversion corresponds to divergent $\theta_B$ and $\theta_M$. The corresponding value of temperature can be read on the right vertical scale. All tasks are in this case obtained.}
\label{temp1}
\end{figure}

Note that the temperature $\theta_B$ of the body increases monotonically with the ratio $p_e/p_g$. By tracing out the machine degrees of freedom from the master equation \eqref{METQ}, one obtains a diagonal state with stationary populations $p_g$ and $p_e$ of the body $B$.
Be now $\dot{Q}_r^B$ the flux $\dot{Q}_r$ seen by $B$. Then the expressions for heat fluxes exchanged by $B$ with its surroundings are
\begin{eqnarray}
\dot{Q}_r^B&=&-\frac{i}{\hbar}\big\langle[H_B,H_{MB}]\big\rangle,\label{dotQRQ}\\
\dot{Q}_d&=&\Gamma_{d}^+\Big[\big\langle\sigma^{\dag}H_B\kappa_{r}\big\rangle-\frac{1}{2}\big\langle\{H_B,\sigma^{\dag}\kappa_{r}\}\big\rangle\Big]\nonumber\\
&+&\Gamma_{d}^-\Big[\big\langle\sigma H_B\kappa_{r}^{\dag}\big\rangle-\frac{1}{2}\big\langle\{H_B,\sigma\kappa_{r}^{\dag}\}\big\rangle\Big]+c.c.\label{QQTQ},\\
\Q_B&=&\Gamma_B^+\Big[\big\langle \sigma^{\dag} H_B \sigma \big\rangle-\frac{1}{2}\big\langle \{ H_B,\sigma^{\dag}\sigma \} \big\rangle\Big]\nonumber \\
&+&\Gamma_B^-\Big[\big\langle \sigma H_B \sigma^{\dag} \big\rangle-\frac{1}{2}\big\langle \{ H_B,\sigma\sigma^{\dag} \} \big\rangle\Big],\label{QB}
\end{eqnarray}
where the mean values are evaluated over the stationary state of the total system. Exploiting its general form \eqref{rhoX}, it is just a matter of straightforward calculations to evaluate all the mean values above.
Imposing the sum of \eqref{dotQRQ}, \eqref{QQTQ} and \eqref{QB} to vanish (first law for $B$, analogous to Eq. \eqref{1stlaw}), one obtains
\begin{equation}\label{popobj}
\frac{p_e}{p_g}=\frac{\Gamma_B^--\Delta(\omega_B)}{\Gamma_B^++\Delta(\omega_B)},
\end{equation}
where
\begin{equation}\label{delta}
\Delta(\omega_B)=2\Lambda(\omega_B)\mathrm{Im}\{c_r\}+\mathrm{Re}\Big\{c_r \big[\Gamma_d^+-\left(\Gamma_d^-\right)^*\big]\Big\}.
\end{equation}
Note now that, thanks to Eq. \eqref{rhoX}, $\langle \sigma\kappa_r^{\dag} \rangle=c_r$ and $\langle \sigma^{\dag}\kappa_r \rangle_{-}=\mathrm{Im}(c_r)$, such that the first term in $\Delta(\omega_B)$ stems from the resonant heat $-\Q_r=-2\hbar\omega_B\Lambda(\omega_B)\mathrm{Im}(c_r)$ exchanged with the machine, while the second is due to the non-local heat flux $\Q_d$.
Eqs. \eqref{popobj} and \eqref{delta} show that the thermal machine works \textit{only if a stationary quantum coherence $c_r$ is present}. Remarkably, it can be shown \cite{Spehner2014} that quantum discord \cite{Ollivier2001} (a key measure of purely quantum correlations) is a monotonic function of the absolute value of the coherence $c_r$ in our system. Differently from previous studies \cite{Correa2013}, here discord between $M$ and $B$ is a necessary condition for any thermodynamic task, and represents a resource the machine can use through the two different processes $\Q_r$ and $\Q_d$. Eq. \eqref{popobj} means that a quantum coherence between machine and body modifies the stationary temperature of the body with respect to $T_B$. This modification is reported in Fig. \ref{temp1}, where the behaviour of $\theta_B$ as a function of the slab-atoms distance $z$ is shown for two different slab thicknesses $\delta$. Four possible regimes can be singled out: both during refrigeration ($\theta_B<T_B$) and heating ($\theta_B>T_B$), $\theta_B$ can be either driven outside of the range $[T_W,T_S]$ (strong tasks) or kept within it (light tasks). As a limiting case of strong heating, the body can be brought to infinite temperature ($p_e=p_g$) and, further on, to negative ones, producing population inversion.
As one can easily see from Fig. \ref{temp1}, the physics behind the absorption tasks is enclosed in the strong sensitivity of the population temperature $\theta_B$ of the body to the population temperature $\theta_M$ of the machine along the resonant transition when the body is not present.
\subsection{Optimal conditions for thermodynamic tasks}\label{optimal}
\begin{figure}[t!]
\begin{center}
\includegraphics[width=240pt]{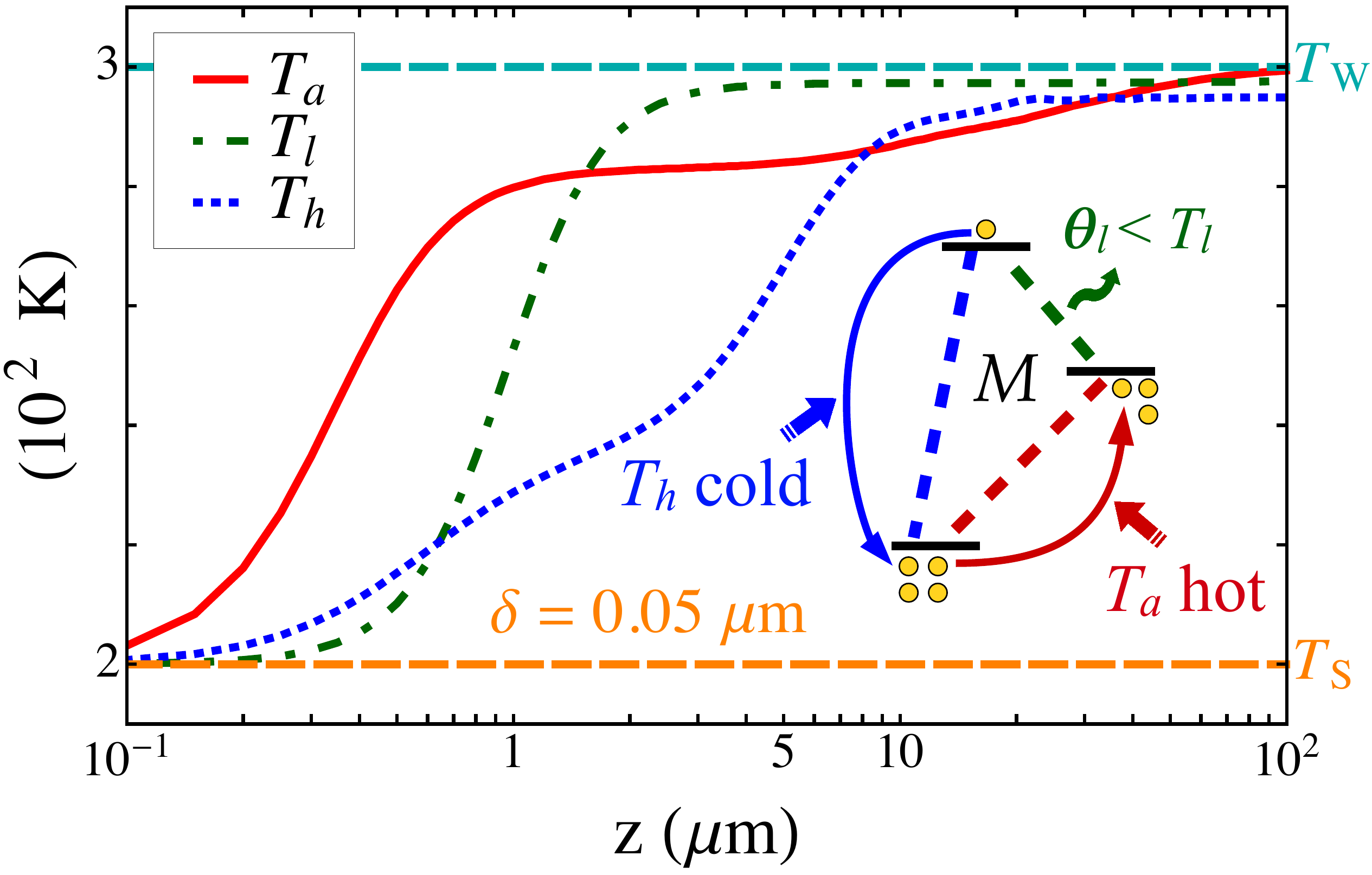}
\end{center}
\caption{Conditions for refrigeration: effective rate temperatures $T_h$, $T_a$ and $T_l$ of the local environments felt by the three machine transitions, for $\omega_h=\omega_S$, $\omega_a=0.8\,\omega_S$ and $\omega_l=0.2\,\omega_S$ ($\omega_S=0.81\cdot 10^{14}$\,rad\,\,$\mathrm{s}^{-1}$) versus the machine-slab distance $z$ in the absence of the body. The slab and walls temperatures are $T_S=200\,$K and $T_W=300\,$K, and the slab thickness is $\delta=0.05\,\mu$m. As the plot shows, the transition having the same frequency as the slab resonance is much more strongly affected by the field emitted by the slab, such that its rate temperature is kept much lower than $T_a$. This produces a mechanism, shown in the inset, according to which excitations (yellow dots) are transferred to the intermediate level of the machine and removed from its upper one. This in turn drives the population temperature $\theta_l$ of the transition at $\omega_l$ (transition 2 in the example) to values lower than the one $T_l$ of its local environment, allowing the machine to refrigerate objects. In this configuration, introducing a body $B$ at $z=1\,\mu$m from the slab and $r=1\,\mu$m from $M$, one obtains $\theta_B=160\,$K $<T_S$.}
\label{temprate}
\end{figure}

It is shown in Fig. \ref{temp1} that the machine has a very high thermal inertia, such that the body, when put into thermal contact with the machine having a certain temperature $\theta_M$, thermalizes with it and $\theta_B\simeq\theta_M$. Fig. \ref{temprate} shows the mechanism the machine uses to modify its population temperature $\theta_M$ in absence of the body, thanks to the different environmental temperatures each of its transition feels. This drives $M$ out of detailed balance condition and allows $M$ to keep its resonant transition temperature almost constant.
We label here the three transitions of the machine as high frequency ($\omega_h$), average frequency ($\omega_a$) and low frequency ($\omega_l$), one of which (suppose here $\omega_l$, connecting states $|0_r\rangle$ and $|1_r\rangle$) is resonant with $\omega_B$. For simplicity, let us focus on refrigeration only, which we suppose to happen either through transition 2 (connecting first and the second excited states), since in this configuration the high-frequency transition $3$ is always used by an absorption refrigerator to dissipate heat into the environment \cite{Correa2014}. As shown in Fig. \ref{temp1}, to obtain a low $\theta_B$ the resonant machine transition must be made cold. This is achieved by reducing the ratio $p_{1_r}/p_{0_r}$, which in turn happens when: \\ (a) the effective environmental temperature $T_h$ felt by the high frequency transition is very cold. In this way the environment contributes in increasing the population of the ground state of $M$ at the expenses of the population of its most energetic state. The resonant transition involves necessarily one of these two levels, and in both cases the effect of the high frequency transition helps reducing $p_{1_r}/p_{0_r}$;\\
(b) the effective environmental temperature felt by the average transition is very hot. This, following the same idea, would either mean reducing the population of $p_1$ or increasing the one of $p_0$, thus reducing $p_{1_r}/p_{0_r}$.\\
When these two conditions are met, the machine can always redistribute its populations such that the ratio $p_{1_r}/p_{0_r}$ can be kept low and almost unaffected by the presence of another atom.
The advantage of the OTE field configuration is that the effective field temperatures can be manipulated through a wide set of parameters involving $z$, $\delta$, $T_W$ and $T_S$. In particular, the role of the resonance of the slab material is crucial \cite{Bellomo2012}, as explained in the caption of Fig. \ref{temprate}. In the case $T_S<T_W$, transitions strongly affected by the field emitted by the slab feel a cold local environment. Moreover, provided  $\omega_a$ is far enough from $\omega_S$, one can at the same time have $T_a\simeq T_W$. By this mechanism, $M$ can change the temperature $\theta_B$, bringing it to values far outside the range $[T_S,T_W]$.

\begin{figure}[t!]
\begin{center}
\includegraphics[width=200pt]{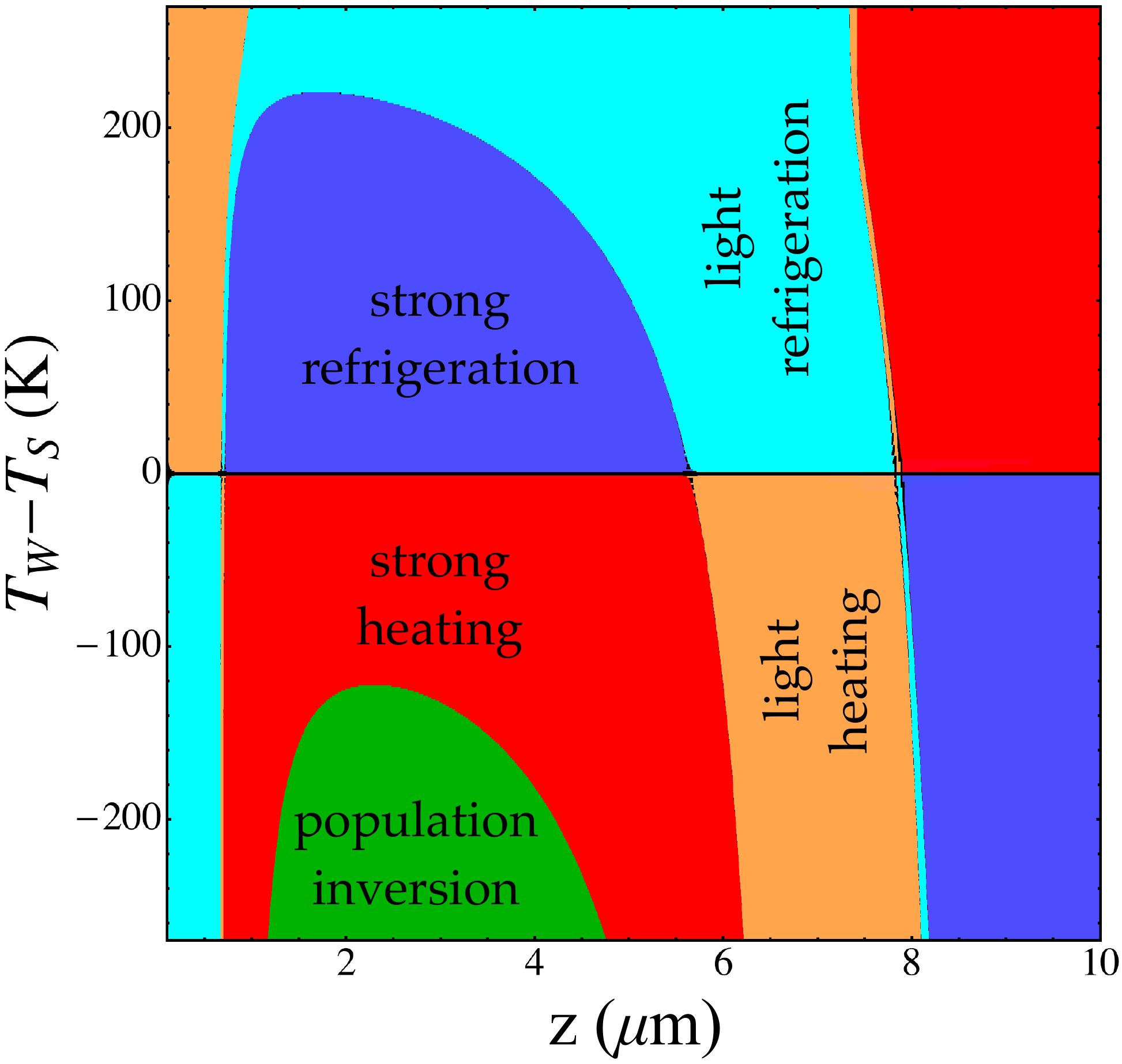}
\end{center}
\caption{Functioning phases of the absorption machine versus the atoms-slab distance $z$ and $\Delta T=T_W-T_S$. The sapphire slab has a thickness $\delta=0.05\,\mu\mathrm{m}$ and its temperature is continuously changed in the range $[30\,\mathrm{K},570\,\mathrm{K}]$. The walls are at $T_W=300\,\mathrm{K}$. Strong refrigeration (blue areas), strong heating (red areas), light refrigeration (cyan), light heating (orange) and population inversion (green) can all be obtained.}
\label{phases}
\end{figure}

We stress here that the difference between light and strong tasks is a fundamental one: better than light tasks could in principle be done by direct connection of the body to one of the two real reservoirs at $T_S$ or $T_W$, while strong tasks can not be achieved by a simple thermal contact with anything in the system.\linebreak
$\Delta(\omega_B)$ strongly depends on the slab-matter system distance $z$ and on the external temperatures through $c_r$, $\Lambda$ and $\Gamma_d^{\pm}$. One can thus engineer one or many of these regimes at will as shown in the functioning-phase diagram of the machine in Fig. \ref{phases} for a fixed thickness $\delta=0.05\,\mu\mathrm{m}$. All the strong and light functioning phases of the machine are found as a function of both $T_W-T_S$ and $z$.

\section{Efficiency and Carnot limit}\label{efficiency} Consider now the refrigerating regime in which the machine extracts heat from the body through the transition $2$. The scheme of heat fluxes is then exactly the one depicted in Fig. \ref{scheme}. The efficiency of this process is
\begin{equation}\label{eff}
\eta_{ref}=\frac{\Q_r}{\Q_1+\Q_2},
\end{equation}
due to the fact that $\Q_r$ is the power produced by the machine, which absorbs energy from its surroundings through transitions $1$ and $2$ (the equivalent of a work input) while uses transition $3$ to dissipate part of the absorbed energy after use (the equivalent of the spiral in a normal fridge). The corresponding Carnot limit $\eta^C_{ref}$ can be obtained by analysing the machine functioning in its reversible limit (zero entropy production).
The instantaneous entropy production rate $\tau$ for quantum systems is defined as \cite{BreuerBook}
\begin{equation}\label{secondlaw}
\sigma=-\frac{\mathrm{d}}{\mathrm{d}t}S(\rho(t)||\rho^{st})\geq0,
\end{equation}
where $S(\rho(t)||\rho^{st})$ is the so-called relative entropy \cite{Vedral2002}, never increasing in time under a Markovian dynamics. Following \cite{Correa2014}, one can apply equation \eqref{secondlaw} term by term to each dissipator in the master equation thanks to the fact that they all are under a Markovian form. One thus obtains
\begin{equation}\label{secondlawlocal}
\sum_i\mathrm{Tr}\Big[(D_k\rho^{st})\ln \rho_k^{st}\Big]\geq 0,
\end{equation}
where $\rho_k^{st}$, $k=1,2,3,B,d$ is the kernel (stationary state) of the single dissipator $D_k$.
The $3$ local dissipators $D_n$ of the machine and the local dissipator $D_B$ of the body induce stationarity under the standard Gibbs form at the effective environmental temperature, diagonal in the free atomic Hamiltonian basis. The nonlocal dissipator $D_d$, in the case studied here where the dipoles of $B$ and $M$ lie along the line connecting the two atoms (and more in general when $\Gamma_d^{\pm}\in \mathbb{R}$), has the same kernel at environmental temperature $T_d$, local in the degrees of freedom of $M$ and the $B$.
Introducing these single-dissipator stationary states into equation \eqref{secondlawlocal} one obtains
\begin{equation}\label{2law}
\frac{\dot{Q}_1}{T_1}+\frac{\dot{Q}_2}{T_2}+\frac{\dot{Q}_3}{T_3}+\frac{\dot{Q}_B}{T_B}+\frac{2\dot{Q}_d}{T_d}\leq 0,
\end{equation}
which is a form of the second law at stationarity for our system.
With the help of the first law in Eq. \eqref{1stlaw} of Section \ref{thermodynamics}, the known property of three-level atomic heat fluxes \cite{Scovil1959} $|\Q_n/\Q_m|=\omega_n/\omega_m$ $\forall \,m,n=1,2,3$ (where $\Q_n$ is the total flux along the $n$-th transition) and the fact that in refrigeration $T_3<T_2,T_d$ and $T_2,T_d<T_1$ (as commented in Section \ref{optimal}) and under the condition $\Q_d<0$ (other cases can be treated analogously), one obtains from \eqref{2law} another first degree inequality. This has a non-trivial solution only if $T_d>T_2$, from which a bound on the efficiency in Eq. \eqref{eff} can be obtained as shown in Appendix B.
Such a bound depends only on the three frequencies of the machine and the temperatures of the effective local and non-local environments. In the case of refrigeration along transition $2$ the Carnot efficiency assumes the form
\begin{equation}\label{carnot}
\eta^C_{ref}=
\begin{cases}
\frac{\omega_2}{2\omega_1}+\frac{1}{2}\frac{T_2T_d(T_1-T_3)}{T_1T_3(T_d-T_2)}+\frac{\omega_2}{2\omega_1}\frac{T_2(T_d-T_3)}{T_3(T_d-T_2)},\,\,\mathrm{if}\,\,T_d>T_2,\\
\frac{\omega_2}{\omega_1},\,\,\mathrm{if}\,\,T_d\leq T_2.
\end{cases}
\end{equation}
\subsection{Efficiency at maximum power}
\begin{figure}[t!]
\begin{center}
\includegraphics[width=240pt]{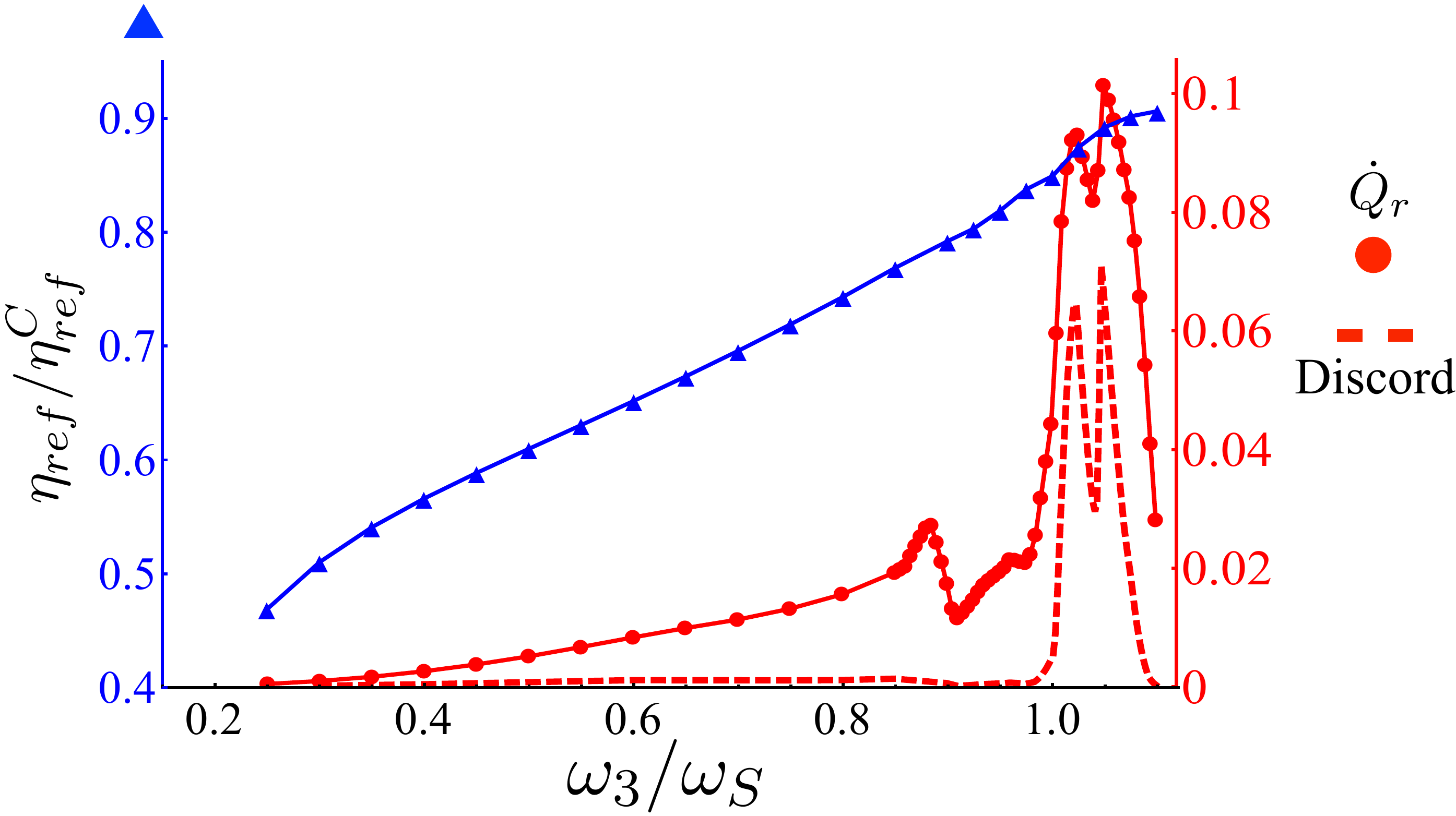}
\end{center}
\caption{The ratio $\eta_{ref}/\eta^C_{ref}$ (blue triangles, left vertical axis), the power of the machine ($\Q_r$, red dots, right vertical axis in units of $10^{-14}\,\mu \mathrm{J}/s$) and machine-body discord (dashed red line, right vertical scale in units of $10^{-4}$) versus the scaled machine transition frequency $\omega_3/\omega_S$. The sapphire slab is semi-infinite and at $T_S=395\,\mathrm{K}$, while $T_W=125\,\mathrm{K}$ and $z=4.8\,\mu\mathrm{m}$. The transition frequency of the body is fixed as $0.1 \omega_S$ ($\omega_S=0.81\cdot10^{14}\,\mathrm{rad}s^{-1}$), resonant with transition $2$ of the machine. The maximum power is reached for $\omega_3=1.05\,\omega_S$, corresponding to $\eta_{ref}/\eta^C_{ref}\simeq 0.89$. Remarkably, discord shows a sharp resonance peak, similarly to $\Q_r$.}
\label{effmax1}
\end{figure}

An important figure of merit for the realistic functioning of any thermal machine is how close to its Carnot limit it works when delivering maximum power (i.e., when $\Q_r$ is maximised). Many bounds are known for different setups, limiting the efficiency at maximum power $\eta^m$ to some fractions of $\eta^C$ \cite{Correa2014, Curzon1975}. Remarkably our structured OTE environment allows for refrigeration tasks with $\eta^m$ much closer to $\eta^C$ than the bound known for quantum absorption machines \cite{Correa2014} based on ideal blackbody reservoirs, reading for our system $\eta^m<0.75\,\eta^C$. This is exemplified in Fig. \ref{effmax1} for a particular configuration of the model. The blue triangles (left vertical scale) represent the ratio $\eta_{ref}/\eta^C_{ref}$, plotted versus $\omega_3$ while keeping fixed $\omega_2=\omega_B=0.1\,\omega_S$. The red dots (right vertical scale) are the power $\Q_r$ plotted versus the same quantity, while the red dashed line is the machine-body discord (right vertical scale). It is clear that the power is maximised at $\omega_3=1.05\,\omega_S$, corresponding to $\eta_{ref}^m\simeq0.89\,\eta^C_{ref}$. $\Q_r$ starts decreasing, as classically expected when the efficiency approaches $\eta^C$, around $\omega_3\simeq0.9\,\omega_S$, but suddenly increases again when $\omega_3$ approaches $\omega_S$. This behaviour is due to the fact that, when one atomic transition is resonant with the characteristic frequency of the slab material, the atomic populations are strongly affected by the field emitted by the slab. Hence the not-black-body nature of the total field become crucial (e.g., the atomic decay rate is no longer proportional to $\omega^3$), allowing to overcome bounds set by the blackbody physics. The role of discord as machine resource is clearly shown here, where discord at resonance has a sharp peak leading to the high-power performance of $M$.

\begin{figure}[t!]
\begin{center}
\includegraphics[width=220pt]{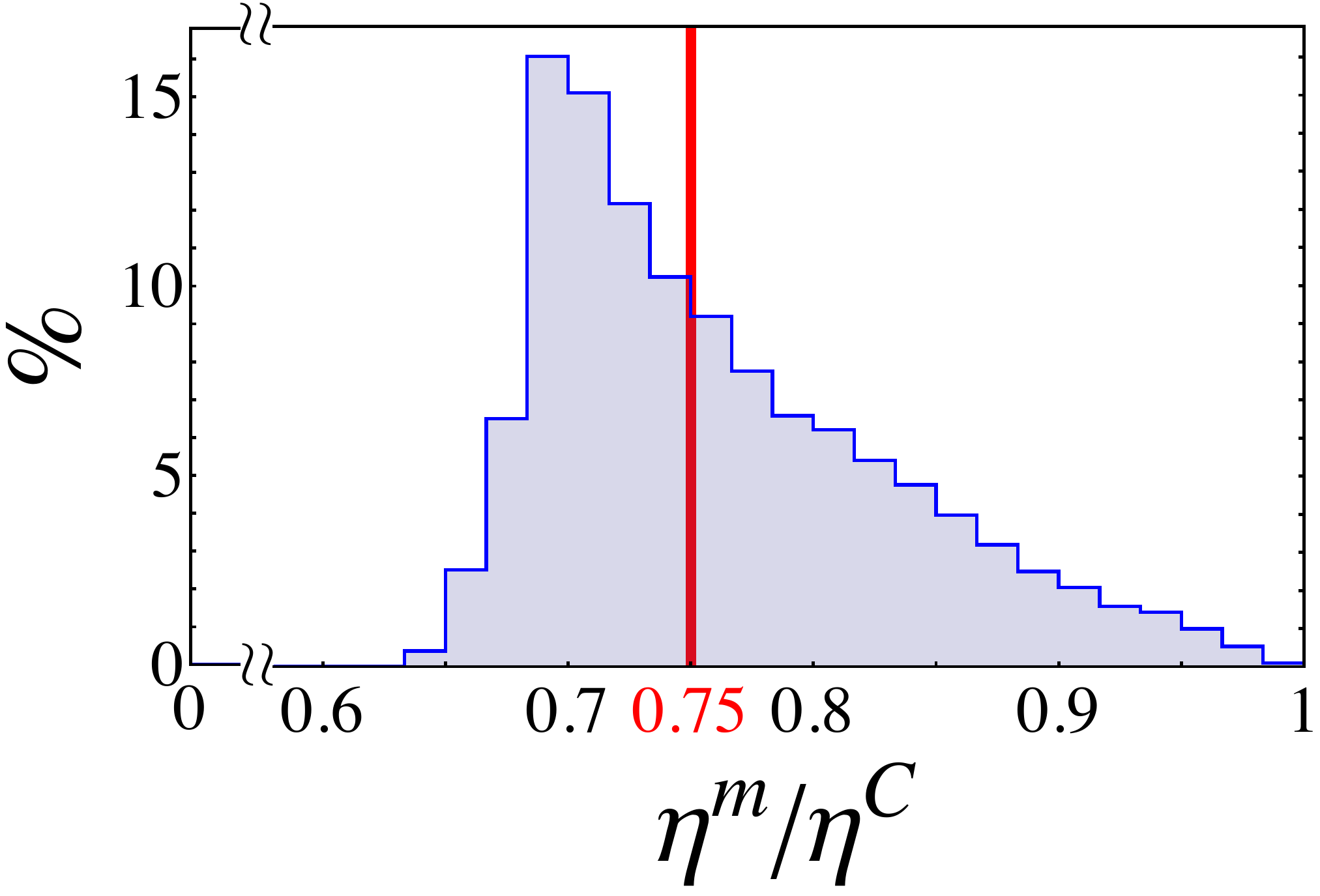}
\end{center}
\caption{Statistical occurrence of ratios $\eta^m/\eta^C$ for a random sampling of $2\cdot10^4$ thermal machines, always in resonance with the same body $B$. For each machine, $z$ has been randomly generated in the range $[0.9\,\mu \mathrm{m}, 100\,\mu \mathrm{m}]$, $T_W\in[50\,\mathrm{K}, 500\,\mathrm{K}]$ and, for each value of $T_W$, $T_S\in [T_W,T_W+500\,\mathrm{K}]$. The internal structure of the body is kept fixed during the simulations, with a frequency $\omega_B=0.1 \,\omega_S$ resonant with the transition 2 of $M$. The maximisation is performed over every possible value of $\omega_1 \in (\omega_2,\omega_S)$ and $\omega_3$ compatible with the condition $\omega_1+\omega_2=\omega_3$. Around $50\%$ of machines thus generated have $\eta^m>0.75\,\eta^C$.}
\label{effmax2}
\end{figure}

One could wonder whether such an exceptionally high efficiency at maximum power is only seldomly attained for the kind of machines described here. To answer such a question on quantitative bases, we performed a random sampling of over $2\cdot10^4$ thermal machines, all delivering thermodynamic tasks on the same fixed body. In the simulations performed and reported in Fig. \ref{effmax2}, the machines work as a quantum refrigerator delivering strong refrigeration using a semi-infinite slab. In this sampling, the machine-slab distance $z$ has been, for each machine, randomly drawn in the range $[0.9\,\mu \mathrm{m}, 100\,\mu \mathrm{m}]$, the walls temperature has been selected randomly in $T_W\in[50\,\mathrm{K}, 500\,\mathrm{K}]$ and, for each value of $T_W$, the slab temperature has been chosen at random in $T_S\in [T_W,T_W+500\,\mathrm{K}]$. The internal structure of the body is kept fixed during the simulations, with a frequency $\omega_B=0.1 \,\omega_S$ resonant with the transition 2 of $M$. For each machine thus generated, we have then maximised the delivered power by modifying the two other machine frequencies over every possible value of $\omega_1 \in (\omega_2,\omega_S)$ and $\omega_3$ compatible with the condition $\omega_1+\omega_2=\omega_3$. Finally, once obtained the configuration corresponding to the maximum power, we have computed the efficiency of the process. Fig. \ref{effmax2} shows the histogram of the distribution of the ratio $\eta^m/\eta^C$ of efficiency at maximum power to the corresponding Carnot efficiency in the interval $[0,1]$ within these $2\cdot10^4$ random refrigerators. It is remarkable that around $50\%$ of these machines work at maximum power with efficiencies higher than the bound $0.75\,\eta^C$ in \cite{Correa2014} and that none of them have been found to work at maximum power with efficiencies lower than $0.6 \eta^C$. Moreover, as can be clearly seen in Fig. \ref{effmax2}, a small but non-negligible fraction of them can reach $\eta^m\simeq0.98\eta^C$.

\section{Conclusions}\label{conclusions} This work introduces a new realization of a quantum thermal machine using atoms interacting with single non-equilibrium electromagnetic fields.  By simply connecting two thermal reservoirs to \textit{macroscopic objects}, their radiated field allows the atomic machine to achieve all quantum thermodynamic effects (heating, cooling, population inversion), without any direct external manipulation of atomic interactions. This overcomes the usual difficulty of connecting single transitions to thermal reservoirs, in a realistic and simple configuration where the field-mediated atomic interaction modifies at will stationary inter-atomic energy fluxes.

Despite the environmental dissipative effects, atoms share steady quantum correlations \cite{Bellomo2013, Bellomo2014}  which we showed to be
necessary for one atom to deliver a thermodynamic task on the other, uncovering genuinely non-classical machine functioning.
These particular features affect the tasks efficiency, which can be remarkably high also at maximum power, defying the known bounds for quantum machines based on ideal and independent blackbody reservoirs thanks to the fundamental effect of the resonance with the real material of which the slab is made. Moreover, such a remarkably high efficiency at maximum power is strongly connected to the presence of a peak in quantum correlations between the machine and the body, which represent the resource the machine uses for its tasks.

These results tackle major open problems on quantum thermal machines, paving the way for an efficient quantum energy management based on the potentialities of non-equilibrium and quantum features in atomic-scale thermodynamics.\\
\section*{Acknowledgments}
The authors acknowledge fruitful discussions with N. Bartolo and R. Messina, and financial support from the Julian Schwinger Foundation.

\begin{appendix}
\section{Resonant heat flux}
In this appendix we demonstrate that the resonant energy exchange between $M$ and $B$ due to the field-mediated coherent interaction $H_{MB}$ consists only of heat.
Following the approach of \cite{Weimer2008}, the dynamics of the sole $M$ induced by the Hamiltonian interaction $H_{MB}$ comprises in general an Hamiltonian and a dissipative part and can be written as
\begin{equation}\label{dotrhoa}
\dot{\rho}_M=-\frac{i}{\hbar}\big[H_M+H_M^{\mathrm{eff}},\rho_M\big]+D_{MB}(\rho),
\end{equation}
where $D_{MB}$ is a non-unitary dissipative term for $M$ due to the interaction with $B$, which depends however on the total state $\rho=\rho_{MB}$ because, in general, the two subparts are correlated. $H_M^{\mathrm{eff}}$ is a renormalised free Hamiltonian of subsystem $M$ due to the interaction with $B$.
Defining the two marginals $\rho_{M(B)}=\mathrm{Tr}_{B(M)}\rho$ and the correlation operator $C_{MB}=\rho-\rho_M\otimes\rho_B$, it is shown in \cite{Weimer2008} that
\begin{eqnarray}
H_M^{\mathrm{eff}}&=&\mathrm{Tr}_B\Big(H_{MB}(\mathbb{I}_M\otimes\rho_B)\Big),\label{HAeff}\\
D_{MB}(\rho)&=&-i \mathrm{Tr}_B\Big([H_{MB},C_{MB}]\Big).\label{DAB}
\end{eqnarray}
Introducing $H_{M1}^{\mathrm{eff}}$ as the part of $H_M^{\mathrm{eff}}$ which commutes with $H_M$ and $H_{M2}^{\mathrm{eff}}$ which does not, directly from equation \eqref{dotrhoa} one has, for the internal energy of $M$ $U_M=\mathrm{Tr}\big((H_M+H_M^{\mathrm{eff}}) \rho\big)$,
\begin{equation}\label{udot}\begin{split}
&\dot{U}_M=\mathrm{Tr}_M\Big((H_M+H_{M1}^{\mathrm{eff}})D_{MB}(\rho)\Big)+\mathrm{Tr}_M\big(\dot{H}_{M1}^{\mathrm{eff}}\rho_M\big)\\
&-i\mathrm{Tr}_M\Big(\big[H_M+H_{M1}^{\mathrm{eff}},H_{M2}^{\mathrm{eff}}\big]\rho_M\Big).
\end{split}\end{equation}
It is custom to identify heat terms as the ones producing a change in the entropy of a subsystem: all the rest is identified as work $W$. Eq. \eqref{udot} can then be split in
\begin{eqnarray}
\Q_M&=&\mathrm{Tr}_M\Big((H_M+H_{M1}^{\mathrm{eff}})D_{MB}(\rho)\Big)\label{dotQA},\\
\dot{W}_M&=&\mathrm{Tr}_M\Big(\dot{H}_{M1}^{\mathrm{eff}}\rho_M-i\big[H_M+H_{M1}^{\mathrm{eff}},H_{M2}^{\mathrm{eff}}\big]\rho_M\Big).\label{dotWA}
\end{eqnarray}
Introducing the symbols $c_M^{ij}=\langle i|\rho_M|j\rangle$ ($i\neq j$) for the coherences \textit{of the marginal} $\rho_M^{st}$ (different then from the coherence $c_r$ introduced in Eq. \eqref{rhoX} of Section \ref{tasks} which is a two-atom coherence), equation \eqref{HAeff} becomes
\begin{equation}
H_M^{\mathrm{eff}}\propto \mathrm{Re}(c_M^{10}).
\end{equation}
By tracing out the machine or the body degrees of freedom from equation \eqref{rhoX}, one can prove that the two stationary marginals $\rho_M^{st}$ and $\rho_B^{st}$ are always diagonal in the eigenbases of their respective free Hamiltonians, so that $c_M^{10}=0$. No renormalisation to the machine Hamiltonian comes therefore from the interaction with $B$, which means that equation \eqref{dotWA} vanishes, proving that no work is involved in machine-body energy exchanges.
As for the heat, considering that $[H_{MB},\rho_M^{st}\otimes\rho_B^{st}]=0$, Eq. \eqref{dotQA} reduces to
\begin{equation}
\Q_M=-i\mathrm{Tr}_M\Big(H_M\mathrm{Tr}_B\big[H_{MB},\rho^{st}\big]\Big)=\Q_r,
\end{equation}
with the same $\Q_r$ given in Eq. \eqref{Qrcorr}.
\section{Carnot limit}
In this appendix we deduce Eq. \eqref{carnot} of Section \ref{efficiency} for the Carnot efficiency in refrigeration along transition 2, and under the condition $\Q_d<0$. In addition to Eqs. \eqref{1stlaw} and \eqref{2law}, the condition $|\Q_n/\Q_m|=\omega_n/\omega_m$ gives for $n=1$ and $m=2$ the following
\begin{equation}\label{ratio12}
\frac{\Q_1}{\Q_2+\Q_r+\Q_d}=\frac{\omega_1}{\omega_2}.
\end{equation}
Solving Eqs. \eqref{1stlaw} and \eqref{ratio12} for $\Q_3$ and $\Q_d$ and using these solutions into \eqref{2law} one obtains for $\Q_r$
\begin{equation}\label{disqr}
\Q_r\leq\Q_1\frac{T_d T_2}{T_d-T_2}\Bigg[\frac{1}{T_3}\Big(1+\frac{\omega_2}{\omega_1}\Big)-\frac{1}{T_1}-\frac{\omega_2}{\omega_1}\frac{1}{T_d}\Bigg]-\Q_2
\end{equation}
which, used in Eq. \eqref{eff} of Section \ref{efficiency}, gives a bound on $\eta_{ref}$ as a function of $\Q_1$ and $\Q_2$. Finally, using the fact that such a bound is a decreasing function of $\Q_2$, one obtains the Carnot efficiency as the limit for $\Q_2\rightarrow 0$, which turns out to be independent on $\Q_1$ and gives ultimately the first line of Eq. \eqref{carnot}.
On the other hand, in the case $T_2<T_d$, one can not obtain anything like Eq. \eqref{disqr} and the only possibility for the machine to work without producing entropy is therefore to have vanishing heat flux from/to the body. This means $\Q_2=\Q_d=0$ which, inserted in the expression for the efficiency and using again $|\Q_n/\Q_m|=\omega_n/\omega_m$ leads to the second line of Eq. \eqref{carnot}.
\end{appendix}

\end{document}